\documentclass{article}
\usepackage{spconf,amsmath,graphicx}
\usepackage{booktabs}

\usepackage{enumitem}
\setlist{nosep, leftmargin=14pt}

\usepackage{mwe} 
\usepackage{pgf}
\usepackage{microtype}
\usepackage{multirow}

\newcommand\blfootnote[1]{%
  \begingroup
  \renewcommand\thefootnote{}\footnote{#1}%
  \addtocounter{footnote}{-1}%
  \endgroup
}

\definecolor{bron}{RGB}{201, 12, 56}
\definecolor{thick}{RGB}{15, 128, 45}
\definecolor{mial}{RGB}{14, 68, 143}
\definecolor{miolar}{RGB}{168, 163, 20}
\definecolor{cons}{RGB}{12, 155, 207}

\begin{document}

\def\x{{\mathbf x}}
\def\L{{\cal L}}

\title{CT evaluation of 2D and 3D holistic deep learning methods for the  volumetric segmentation of airway lesions}

\name{
\begin{tabular}{c}
\textit{Amel Imene Hadj Bouzid}$^1$ \qquad \textit{Baudouin Denis de Senneville}$^2$ \qquad \textit{Fabien Baldacci}$^3$ \\ 
\textit{Pascal Desbarats}$^3$  \qquad \textit{Patrick Berger}$^1$ \qquad \textit{Ilyes Benlala}$^4$ \qquad \textit{Gaël Dournes}$^4$
\end{tabular}
}

\address{$^1$Univ. Bordeaux, INSERM, CRCTB, U 1045, F-33000 Bordeaux, France \\ 
         $^2$Univ. Bordeaux, CNRS, Bordeaux INP, IMB, UMR 5251, F-33400 Talence, France\\
         $^3$Univ. Bordeaux, CNRS, Bordeaux INP, LaBRI, UMR 5800, F-33400 Talence, France \\
         $^4$Service d’Imagerie Médicale Radiologie Diagnostique et Thérapeutique, CHU de Bordeaux, France       
         }

%

%
\maketitle

\begin{abstract}
This research embarked on a comparative exploration of the holistic segmentation capabilities of Convolutional Neural Networks (CNNs) in both 2D and 3D formats, focusing on cystic fibrosis (CF) lesions. The study utilized data from two CF reference centers, covering five major CF structural changes. Initially, it compared the 2D and 3D models, highlighting the 3D model's superior capability in capturing complex features like mucus plugs and consolidations. To improve the 2D model's performance, a loss adapted to fine structures segmentation was implemented and evaluated, significantly enhancing its accuracy, though not surpassing the 3D model's performance. The models underwent further validation through external evaluation against pulmonary function tests (PFTs), confirming the robustness of the findings. Moreover, this study went beyond comparing metrics; it also included comprehensive assessments of the models' interpretability and reliability, providing valuable insights for their clinical application.

\end{abstract}

\blfootnote{© 2024 IEEE International Symposium on Biomedical Imaging.  Personal use of this material is permitted.  Permission from IEEE must be obtained for all other uses, in any current or future media, including reprinting/republishing this material for advertising or promotional purposes, creating new collective works, for resale or redistribution to servers or lists, or reuse of any copyrighted component of this work in other work}
\section{Introduction}
\label{sec:intro}
Deep learning algorithms are increasingly used in semantic segmentation of medical imaging, especially in identifying and segmenting multi-class disease-related abnormalities in CT scans \cite{R2}. This progress is crucial in airway imaging, where non-invasive methods are key for quantifying and characterizing lung abnormalities, essential in tracking the progression and severity of diseases like cystic fibrosis (CF) \cite{R3}.

Cystic fibrosis, characterized by excessive mucus production, results in serious respiratory complications. CT scans play a vital role in evaluating lung disease severity in CF. However, the commonly used visual scoring systems are subjective, with interpretations varying among evaluators \cite{R6}. Moreover, these systems typically categorize lung damage in broad terms, potentially lacking the sensitivity required for detailed, longitudinal monitoring of the disease.

As CF treatments advance, the need for non-invasive, quantitative biomarkers to assess lung disease has become more pressing \cite{R4}. While 2D CNNs have been effective in detecting CF markers in CT scans \cite{R1}, the full potential of 3D volumetric analysis to accurately depict the diverse forms of CF lesions remains largely unexplored. These lesions, varying from tubular bronchiectasis and peribronchial thickening to centrilobular mucus with "tree-in-bud" patterns, and dense consolidations, suggest that a 3D approach could offer more comprehensive insights into these complex structures.

In addressing the segmentation challenges of CF lesions, we employed the nnU-Net architecture in both its 2D and 3D variants for semantic segmentation \cite{R7}. This approach facilitated an in-depth comparison between these methodologies in segmenting CF lesions, which are characterized by a variety of shapes, spatial distributions, and textures. We also refined the nnU-Net's loss function, moving beyond the traditional Dice coefficient to better address the segmentation of both larger and smaller, more intricate lesions \cite{R8}. 
\section{Overview}
\label{sec:met}

In this study, we utilized 2D and 3D nnU-Net networks with customized loss functions to segment CF lesions from 75 clinical exams. For the training phase, we selected 50 CT scans, ensuring a balanced representation of both healthy individuals and patients with a range of lesion counts. For the network input, two modalities were prepared: one including the lung envelope and another excluding it.

\begin{figure}[htb]
\begin{minipage}[a]{1.0\linewidth}
  \centering
  \centerline{\includegraphics[width=8.5cm]{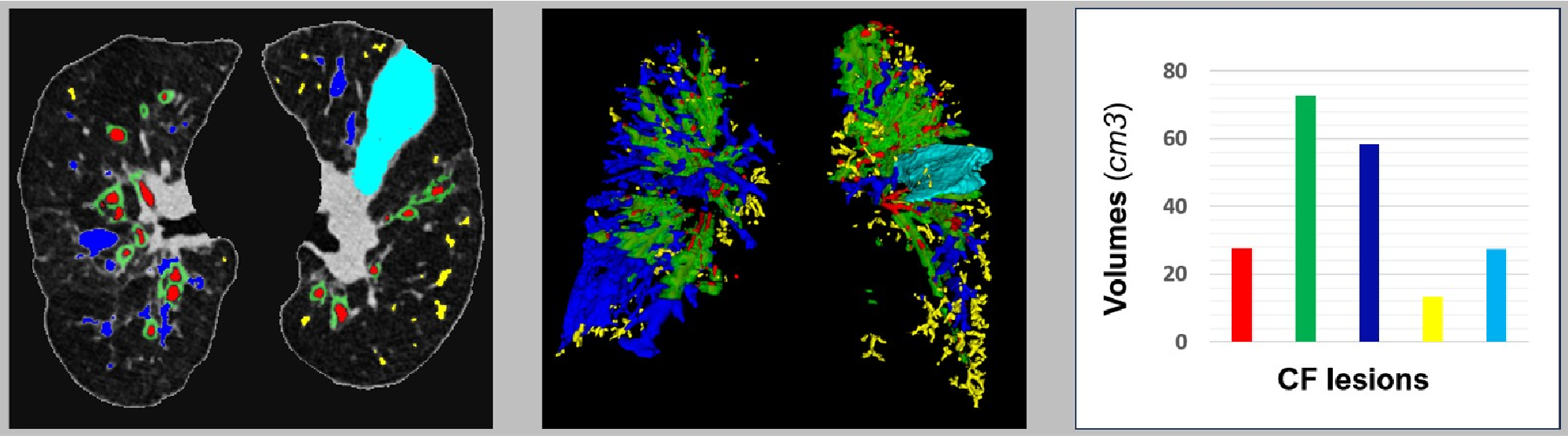}}
  \centerline{(a) Example of patient labeling.}\medskip
\end{minipage}
\begin{minipage}[b]{1.0\linewidth}
  \centering
  \centerline{\includegraphics[width=8.5cm]{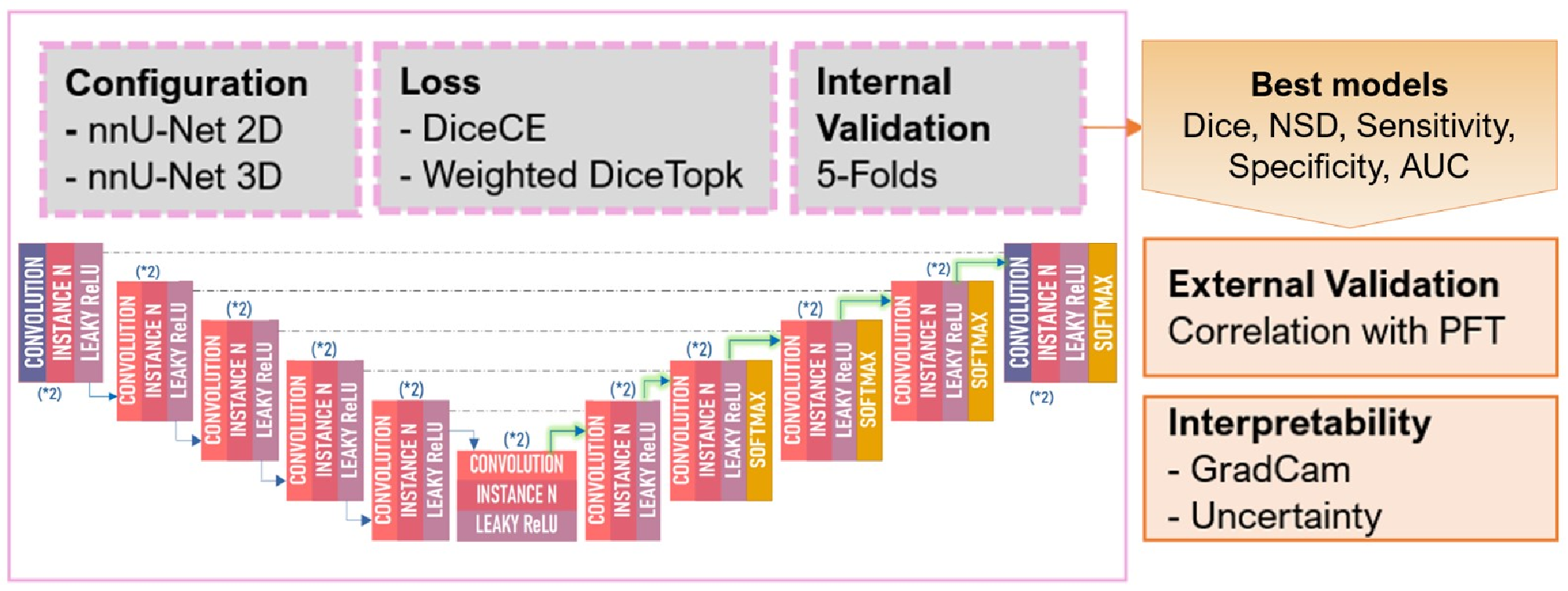}}
  \centerline{(b) Training and Testing Pipeline}\medskip
\end{minipage}
\caption{Study Method. CF lesions are color-coded
as follows: Red for Bronchiectasis, Green for Peribronchial Thickening, Blue for Bronchial Mucus, Yellow for Bronchiolar
Mucus, and Cyan for Consolidation.}
\label{fig:method}
\end{figure}

The internal evaluation involved cross-validation on the 50 CT using metrics specifically designed for detecting lesions of varying sizes. The best-performing models in 2D and 3D were then externally evaluated on a separate cohort of 25 patients, correlating their performance with FEV1\% scores. Additionally, these models underwent model-agnostic analysis with Gradient-weighted Class Activation Mapping (Grad-CAM) \cite{R12} and data-agnostic uncertainty calculation to provide interpretability and enable a comprehensive comparison between 2D and 3D models. The methodological approach is illustrated in the Figure \ref{fig:method}.

\section{Experimental Setup}
\label{sec:met}
 
\subsection{Data Description}
\label{sssec:model}
75 CT scans were conducted using equipment from major manufacturers (GE Revolution® and Siemens Somatom Force®), with slice thicknesses ranging from 1 to 1.25 mm. Manual lesion segmentation was carried out using 3D Slicer 4.11 software by three experienced observers from a CF reference center. The CT images were displayed using parenchymal window settings (width at 1500 Hounsfield Units; level at -450 Hounsfield Units). For validation purposes, a minimum of 80\% visual agreement among the observers was required.

\subsection{Segmentation model}
\label{sec:model}
nnU-Net served as the cornerstone of our segmentation framework. Its encoder, comprising multiple blocks, each contained a sequence of layers. These layers included a convolutional layer, dropout, instance normalization, and Leaky ReLU activation, collaboratively working to extract and process features from the input data.
The decoder in nnU-Net was divided into two main sections. The first section involved deconvolution, crucial for creating high-resolution feature maps, and was constituted by a single convolutional layer. The second section, known as Features Localization, consisted of three layers: a convolutional layer, instance normalization, and Leaky ReLU activation. The convolution kernel sizes in the 2D version of nnU-Net were $3\times3$ with a $2\times2$ stride, while the 3D variant utilized $3\times3\times3$  kernels with a stride of $2\times 2\times2$.
Skip connections between the encoder and decoder established direct links between layers at various depths within the network, facilitating the propagation of information and contributing to enhanced segmentation accuracy. The learning rate (LR) schedule began at 10$^{-2}$ and was updated every 30 epochs. In the optimization process, Stochastic Gradient Descent (SGD) with Nesterov momentum of 0.99 was employed, aiding in iterative parameter updates and model convergence. Notably, nnU-Net used patch sizes of $512\times512$ in 2D and $96\times160\times160$ in 3D, with batch sizes set to 12 in 2D and 2 in 3D.

\subsection{Loss Function}
We considered the \textbf{cross-entropy} (CE), which was defined as \( CE = -\frac{1}{N} \sum_{i=1}^{N} \left[ y_i \log(\hat{y}_i) + (1 - y_i) \log(1 - \hat{y}_i) \right] \), where \( y_i \) was the true label of pixel \( i \), \( \hat{y}_i \) was the predicted probability for that pixel, and \( N \) represented the total number of pixels in the image. Additionally, the \textbf{Dice coefficient} was utilized, defined as \( \text{Dice} = \frac{2 \times \sum_{i=1}^{N} y_i \hat{y}_i}{\sum_{i=1}^{N} y_i + \sum_{i=1}^{N} \hat{y}_i} \), where \( y_i \) and \( \hat{y}_i \) were the true and predicted labels, respectively, for each pixel \( i \).

The initial loss function was defined as \( \text{Loss} = CE + \text{Dice} \), which had been effective in accurately segmenting larger lesions \cite{R10}. However, considering the complexity of tasks involving lesions of various sizes, we adopted a loss function that initially prioritized the Dice coefficient for the effective segmentation of larger lesions within the folds. As training progressed, the focus shifted towards improving the segmentation of the 50\% of pixels that were less accurately predicted by the model. This loss function, which leveraged cross-entropy as its base and incorporated overlap measurements as weighted regularizers, demonstrated stability during training, particularly in adapting to smaller structures \cite{R9}. The Top50 variant of CE, focusing on the most challenging 50\% of the pixels, was defined as \( \text{Top50} = -\frac{1}{N} \sum_{i \in \mathcal{P}_{50}} \left[ y_i \log(\hat{y}_i) + (1 - y_i) \log(1 - \hat{y}_i) \right] \), where \( \mathcal{P}_{50} \) represented the set of pixels with the largest prediction errors.

The customized loss function was the Weighted Dice Top50, formulated as \( \text{WDiceTop50} = (1 - \alpha) \times \text{Dice} + \alpha \times \text{Top50} \), where \( \alpha \) indicated the ratio between the current training epoch and the total number of epochs.

\subsection{Internal Evaluation}
\label{sssec:interpret}
To assess the accuracy of the CF lesion segmentation models, the Dice coefficient was primarily utilized. This measure was proficient in determining the degree of overlap between predicted and actual lesion areas. However, its performance was reduced for smaller structures \cite{R11}. Consequently, the Normalized Surface Distance (NSD) was also considered, owing to its strength in identifying cases where predictions were close but not exactly coincident with the actual lesions \cite{R8}. The NSD assessed the similarity between the predicted segmentation and the ground truth, taking into account a margin around the object's boundary. In our study, this margin was set at 3 pixels (=1.8mm) to accommodate small variations or uncertainties at the segmentation's edge.
Sensitivity and Specificity were used to ensure the thorough detection of significant structures. Furthermore, the Area Under the Curve (AUC) reflected the model's ability to effectively differentiate between various classes.

\subsection{External Evaluation }
\label{sssec:interpret}
In the external evaluation of our study, we specifically focused on the correlation between the volumes predicted by the 2D and 3D nnU-Net models and the FEV1\% volumes measured in a cohort of 25 patients. FEV1\% indicates the volume of air a patient can forcibly exhale in one second.
To establish this relationship, we employed the Spearman's rank correlation test, which was particularly apt for our analysis due to the involvement of non-linear relationships and non-normally distributed data. Our goal was to validate the results of the internal evaluation by confronting the models with clinical data. 

\subsection{Interpretability}
\label{sssec:interpret}
\subsubsection{Model-agnostic}
To enhance our understanding of nnU-Net's functionality, we used Grad-CAM \cite{R12}, which produces heatmaps highlighting key image regions according to the model's focus. This process involved isolating each label, inputting unseen CT-images into nnU-Net, calculating gradients for feature maps in the final convolutional layer, and determining the importance of each feature map. A final Grad-CAM heatmap was generated by applying ReLU activation to emphasize positively contributing features.
\subsubsection{Data-agnostic}
We modified the network into a Bayesian framework, treating weights as probabilistic distributions. This involved using Monte Carlo Dropout \cite{R13} with a 0.3 dropout rate in the convolutional block, creating five distinct network versions. Predictions were computed for each version, and variance among these predictions was calculated across cross-validation folds. This approach quantified the model's uncertainty and provided insights into its prediction confidence.

\section{Results and Discussion}
\label{sssec:results}

\begin{table*}[htb]
\centering
\footnotesize
\begin{tabular*}{\textwidth}{@{\extracolsep{\fill}} cccccccc}
\toprule
Metrics & Configuration & \shortstack{\textcolor{bron}{Bronchiectasis}} & \shortstack{\textcolor{thick}{Peribronchial}\\\textcolor{thick}{Thickening}} & \shortstack{\textcolor{mial}{Bronchial}\\\textcolor{mial}{mucus}} & \shortstack{\textcolor{miolar}{Bronchiolar}\\\textcolor{miolar}{mucus}} & \shortstack{\textcolor{cons}{Consolidation}} & Avg\\
\midrule
 	
 & 2D \textit{DiceCE} & 0.77  & 0.63 & 0.55 & 0.28 & 0.61  & 0.57 \\
Dice  & 2D \textit{WDiceTop50} & \textbf{0.80}  & 0.64  & 0.61  & 0.32 & 0.59 & 0.59 \\
 & 3D \textit{DiceCE}  & \textbf{0.80} & 0.64  & \textbf{0.64} & \textbf{0.39} & \textbf{0.61}  & 0.61 \\
 & 3D \textit{WDiceTop50} & \textbf{0.80} & \textbf{0.65} & \textbf{0.64} & \textbf{0.39} & 0.60 & \textbf{0.62} \\
  	    	  
\hline
 & 2D \textit{DiceCE} & 0.79 & \textbf{0.81} & 0.60  & 0.44  & 0.53  & 0.63 \\
NSD  & 2D \textit{WDiceTop50} & 0.78 & 0.79 & 0.62 & 0.46 & 0.59  & 0.65 \\
 & 3D \textit{DiceCE}  & \textbf{0.80}  & 0.80 & 0.67 & \textbf{0.51} & \textbf{0.67} & 0.69 \\
 & 3D \textit{WDiceTop50} & \textbf{0.80}  & \textbf{0.81}  & \textbf{0.68}  & \textbf{0.51}  & 0.65 & \textbf{0.70} \\

\hline
 & 2D \textit{DiceCE} & 0.74 & 0.63  & 0.47  & 0.22  & \textbf{0.65}  & 0.54 \\
Sensibility  & 2D \textit{WDiceTop50} & 0.76 & \textbf{0.64}  & 0.52  & \textbf{0.43}  & 0.62  & 0.59 \\
 & 3D \textit{DiceCE}  & \textbf{0.77}  & 0.63 & \textbf{0.58}  & 0.34 & 0.64  & 0.59 \\
 & 3D \textit{WDiceTop50} & \textbf{0.77} & \textbf{0.64}  & 0.57  & 0.36  & 0.61 & \textbf{0.60} \\
  
\hline
 & 2D \textit{DiceCE} & 1.00  & 1.00 & 1.00  &\textbf{ 1.00} & 0.99 & \textbf{0.99} \\
Specificity  & 2D \textit{WDiceTop50} & 1.00  & 1.00 & 1.00  & 0.80 & 0.99  & 0.96 \\
 & 3D \textit{DiceCE}  & 1.00  & 1.00  & 1.00  & \textbf{1.00}   & 0.99  & \textbf{0.99} \\
 & 3D \textit{WDiceTop50} & 1.00  & 1.00  & 1.00  & \textbf{1.00}  & 0.99 & \textbf{0.99} \\
 
\hline
 & 2D \textit{DiceCE} & 0.87  & 0.81  & 0.73  & 0.61  & \textbf{0.82}  & 0.77 \\
AUC  & 2D \textit{WDiceTop50} & \textbf{0.88}  & \textbf{0.82}  & 0.76  & 0.61  & 0.80  & 0.77 \\
 & 3D \textit{DiceCE}  & \textbf{0.88} & 0.81  & \textbf{0.79}  & 0.67 & \textbf{0.82} & 0.79 \\
 & 3D \textit{WDiceTop50} & \textbf{0.88} & \textbf{0.82} & 0.78 & \textbf{0.68}  & 0.80 & \textbf{0.80} \\
\bottomrule
\end{tabular*}
\caption{Internal Evaluation of the nnU-Nets predictions. Best results are reported with bold characters.}\label{intern}
\setlength{\tabcolsep}{5.6pt}
\end{table*}

\begin{table}[htb]
\centering
\small
\begin{tabular}{cccc}
\toprule
n=25 &  & \shortstack{2D\\FEV1\%} & \shortstack{3D\\FEV1\%} \\
\midrule
\textcolor{bron}{Bronchiectasis} & $\rho$ & -0.46 & -0.60 \\
& p-value &  \textbf{0.02} &\textbf{ $<$0.001}\\
\hline
\textcolor{thick}{Peribronchial Thickening} & $\rho$ & -0.45 & -0.59 \\
& p-value &  \textbf{0.02} & \textbf{$<$0.001} \\
\hline
\textcolor{mial}{Bronchial mucus} & $\rho$ & -0.34 & -0.50 \\
& p-value &  0.09 & \textbf{0.005} \\
\hline
\textcolor{miolar}{Bronchiolar mucus} & $\rho$ & -0.30 & -0.49 \\
& p-value &  0.10 & \textbf{0.006} \\
\hline
\textcolor{cons}{Consolidation} & $\rho$ & -0.24 & -0.39 \\
& p-value &  0.18 & \textbf{0.04} \\
\bottomrule
\end{tabular}
\caption{External Evaluation on FEV1\% of the modified 2D and 3D nnU-Nets. $\rho$ = Spearman correlation coefficient.}\label{extern}
\end{table}

\begin{figure}
	\centering
		\includegraphics[width=8.5cm]{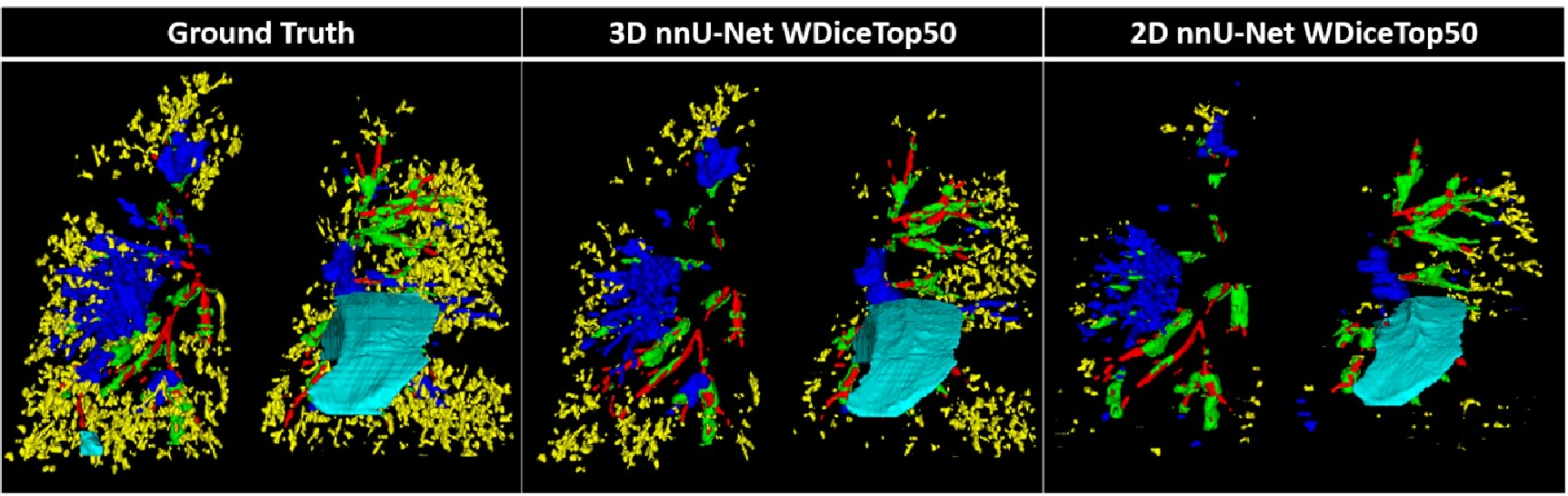}
	\caption{The comparison of CF lesion segmentations is presented. The first column displays the Ground Truth 3D segmentations, while the second and third columns show the predictions by the modified nnU-Net 3D and 2D models, respectively.}
	\label{fig:volumes}
\end{figure}

The internal evaluation, as shown in Table \ref{intern}, revealed that both 2D and 3D nnU-Net models were effective in segmenting bronchiectasis and peribronchial thickening, achieving Dice scores of 80\% and 64\%, respectively, and NSD scores of 80\% for both. The 3D model, however, demonstrated superior performance in detecting smaller lesions such as mucus and consolidations. The custom loss function, designed to enhance the detection of smaller structures, significantly improved the 2D model's performance. Specifically, bronchial and bronchiolar mucus detection improved by 6\% and 4\% (p$<$0.05) in Dice scores, and 2\% (p$<$0.05) in NSD for both. While the Dice score for consolidation detection in the 2D model decreased by 2\%, the NSD increased by 6\%, indicating more accurate detection despite imperfect overlap in multiple slices. Despite these improvements, the 2D model could not match the 3D model's performance, which saw a marginal increase in performance due to the change in loss function, with a  1\% overall increase in Dice and NSD scores. Sensitivity and specificity analysis showed that all four models were specific, with the modified loss enhancing the sensitivity of the 2D model. However, the 2D model's AUC of 77\% was still lower than the 3D model's 80\%.

\begin{figure}
	\centering
		\includegraphics[width=8.5cm]{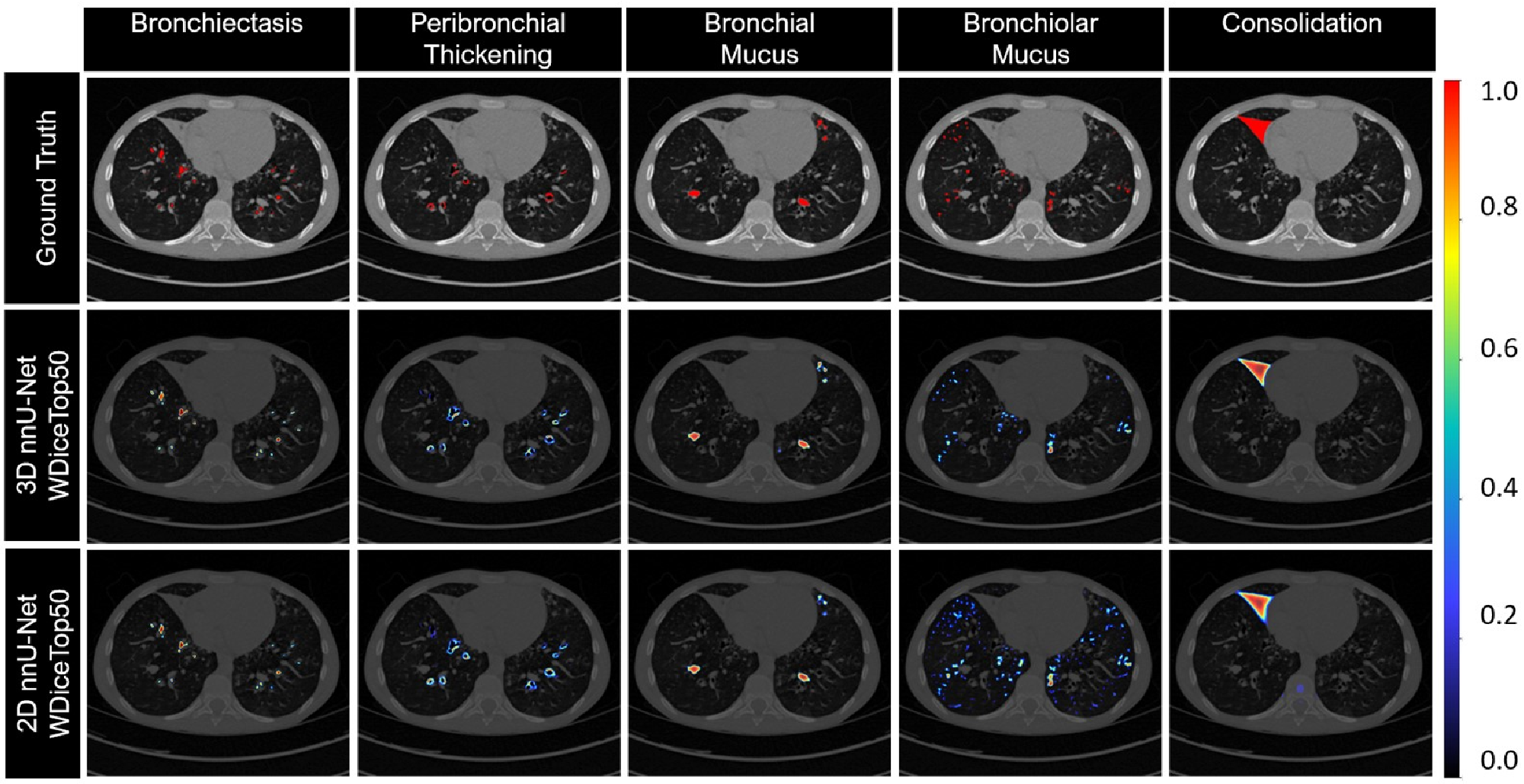}
	\caption{The analysis showcases CF lesion feature maps: Ground Truth segmentations in the first row, followed by GradCAM feature maps of modified nnU-Net 3D and 2D models in the second and third rows, respectively. The maps' intensity levels correlate with prediction probabilities, where higher intensity signifies higher probabilities.}
	\label{fig:gradcam}
\end{figure}

For external evaluation, as detailed in Table \ref{extern}, models with the modified loss were selected based on their superior performance in internal validation. Correlation analysis with FEV1\% indicated significant correlations for bronchiectasis and peribronchial thickening volumes predicted by the modified 2D and 3D nnU-Net models. However, correlations for mucus and consolidation were not significant for the 2D model, whereas they were significant for the 3D model, supporting the findings of the internal evaluation. Figure \ref{fig:volumes} presents an example in 3D of CF predictions, where mucus and consolidation are less detected in the modified nnU-Net 2D compared to its 3D counterpart.

Interpretability, as demonstrated in Figure \ref{fig:gradcam} with GradCAM feature maps for the same patient, showed that the 2D model exhibited more hesitation, particularly in identifying mucus, compared to the 3D model. Additionally, uncertainty estimation, calculated through variances across the five versions of nnU-Net in both 2D and 3D, indicated only a 10$^{-4}$ deviation, suggesting comparable robustness in network predictions between the 2D and 3D models.

Our study challenged the prevalent belief in the literature \cite{R16} about the superiority of 2D models, showing that 3D models perform better in segmenting diverse CF lesion forms, notably tubular bronchiectasis and centrilobular mucus. The 3D models' performance is attributed to their efficient use of spatial information, a crucial factor even with limited training data. The importance of this discovery is amplified by our dataset's challenges, dominated by small mucus plugs and bronchiolar impactions, as depicted in grads \ref{fig:method} and \ref{fig:volumes}.

Exploring the use of separate networks for each lesion type and analyzing metrics individually for each class \cite{R14} could provide more in-depth insights. Furthermore, incorporating attention mechanisms \cite{R17} might enhance decoder guidance through spatial attention maps. This study contributes to the ongoing discourse on the role of Dice loss in semantic segmentation \cite{R15}, highlighting the necessity for a variety of metrics and approaches in the design and evaluation of networks, particularly in the complex context of multi-class segmentation.

\section{Conclusion}
\label{sec:conclusion}
The 2D and 3D models demonstrate comparable effectiveness in detecting bronchiectasis and peribronchial thickening. However, the 3D model is better in identifying more complex lesions like mucus and consolidations, benefiting from its ability to analyze spatial relationships across sequential slices. Our study also highlights the potential advantages of alternative loss functions over the traditional Dice loss in semantic segmentation, particularly for nuanced structures. The results of this study could pave the way for new, fully automated, volumetric scoring systems to quantify CF severity in future clinical applications.

\vfill
\pagebreak

\section{Compliance with ethical standards}
\label{sec:ethics}
All patients were provided written informed consent for reusing data from their medical records according to French Laws and after Institutional Review Board approval (study registration number CHUBX2020RE0267).

\section{Acknowledgments}
\label{sec:acknowledgments}
This study was conducted in the framework of the University of Bordeaux's IdEx "Investments for the Future" program RRI "IMPACT" which received financial support from the French government.

\bibliographystyle{IEEEbib}
\bibliography{2024_ISBI_Imene,refs}

\end{document}